# FERROELECTRIC NANOTUBES

F. D. Morrison,[1] Y. Luo,[2] I. Szafraniak,[2] V. Nagarajan,[3] R. B. Wehrspohn,[2]

M. Steinhart,[2,4] J. H. Wendroff,[4] N. D. Zakharov,[2] E. D. Mishina,[5,6] K. A. Vorotilov,[5]

A. S. Sigov,[5] S. Nakabayashi,[6] M. Alexe,[2] R. Ramesh,[3] and J. F. Scott,[1*]

[1] Symetrix Centre for Ferroics, Earth Sciences Dept., University of Cambridge, Cambridge CB2 3EQ, UK
[2] Max Planck Institute of Microstructure Physics, Weinberg 2, D-06120 Halle, Germany
[3] University of Maryland, College Park, MD 20742, USA
[4] Institute of Physical Chemistry, University of Marburg, Hans-Meerwein, Germany
[5] Moscow State Institute for Radioengineering, Electronics and Automation (Technical University - MIREA), Moscow 119454, Russia
[6] Department of Chemistry, Saitama University, Saitama, 338-8570, Japan

*Correspondence should be addressed to JFS.          email: jsco99@esc.cam.ac.uk

**ABSTRACT:**

We report the independent invention of ferroelectric nanotubes from groups in several countries. Devices have been made with three different materials: lead zirconate-titanate $PbZr_{1-x}Ti_xO_3$ (PZT); barium titanate $BaTiO_3$; and strontium bismuth tantalate $SrBi_2Ta_2O_9$ (SBT). Several different deposition techniques have been used successfully, including misted CSD (chemical solution deposition) and pore wetting. Ferroelectric hysteresis and high optical nonlinearity have been demonstrated. The structures are analyzed via SEM, TEM, XRD, AFM (piezo-mode), and SHG. Applications to trenching in Si dynamic random access memories, ink-jet printers, and photonic devices are discussed. Ferroelectric filled pores as small as 20 nm in diameter have been studied.

Nanotubes of conducting materials such as carbon have recently received considerable attention. Micro-tubes of nonconducting BN and SiC have also been reported in the literature *(1-3)*. However, ferroelectric nanotubes made of oxide insulators, as reported here, have a variety of applications for pyroelectric detectors, piezoelectric ink-jet printers, and memory capacitors that cannot be filled by other nanotubes *(4-8)*. The integration of ferroelectric nano-tubes into Si substrates, shown here, is particularly important in construction of [3D] memory devices beyond the present stacking and trenching designs, which according to the international ULSI schedule *(9)* must be achieved by 2008. Hernandez et al. recently demonstrated the fabrication of ferroelectric nano-tube bundles from a porous alumina substrates; we emphasise that their bundles are not ordered arrays but instead spaghetti-like tangles of nano-tubes that cannot be used for the Si device embodiments considered here *(10)*. Much larger (>20-micron diameter) ferroelectric micro-tubes have been made previously by sputter deposition around polyester fibres *(3,11)* – Glen Fox at EPFL has made them from ZnO and PZT, with 23-$\mu$m inside diameter, about 1000x larger than the smallest nano-tubes reported in the present paper. Application advantages of these devices include *(11)* "enhancing the piezoelectric response of composite materials by five orders of magnitude in comparison with bulk ceramic materials" and for MEMS. A similar process involving coating rather than filling pores, is given by Refs. *1-3,11*, with BN/SiC tubes claimed to be piezoelectric. We report here fabrication of ferroelectric nanotubes of various compositions with structures as small as 20 nm using several techniques. By using porous Si substrates as templates it is also possible to produce regular arrays of discrete nanotubes.

There is a need to deposit uniform coatings of dielectrics on the inside walls of trenches for Gbit Si random-access memories (RAMs), both dynamic DRAMs and non-volatile ferroelectric FRAMs. Thin-film devices, including dielectric capacitors for random-access memories (RAMs) are generally deposited via chemical vapour deposition (CVD), sputtering, or related techniques in which the principal aims are to achieve uniform thickness along the inside walls of a high aspect-ratio trench. At present a typical value for the aspect ratio desired is 6 microns deep by 0.1-micron diameter *(12)*. Here we report such ratios with uniform coatings as thin as 40 nm. These nano-tubes can also be used for ultra-small ink-jet printers or for photonic device arrays. The state-of-the-art ink-jet printer (Seiko-Epson) can deliver a 1.0-picoliter droplet and paint a 1.0-micron line for lithography-free printing of integrated circuits *(13)*; our piezoelectric nanotubes have diameters as small as 0.18 microns, thus suggesting their use for improved nano-ink-jet devices for higher resolution lithography-free integrated circuit production.

In the present work we report successful deposition at such aspect ratios of dielectrics such as lead zirconate titanate PZT, barium titanate, and strontium bismuth tantalate (SBT) for ferroelectric RAMs. Diameters can be as small as 0.1 microns, with 40-200 nm wall thicknesses and very deep depths (> 80 μm). This has important implications for Gbit devices and for three-dimensional [3D] integration.

**Chemical Solution Deposition**

We report here the use of liquid source misted chemical deposition *(14,15)* for filling porous Si substrates. The process involves generation of a fine mist of liquid precursor droplets, of average diameter 0.3 microns. It is particularly important that each droplet is charged, typically at -5e, as determined by prior studies *(16)*. The porous Si substrates used in the process are prepared by photo-excitation of majority and minority carriers, concomitant with a wet etch *(17,18)*. Using this technique, it is possible to produce regular arrays of pores with diameters ranging from a few hundred nanometers to a few microns, with depths of up to 100 μm. Although not emphasized in the literature, the etching process results in a net positive charge for the inside of the Si trenches. Strontium bismuth tantalate was deposited on porous Si substrates using a Samco liquid source misted chemical deposition apparatus.

It is at first surprising that the charged mist droplets produced during the deposition technique do not simply form a miniscus at the top of each pore, particularly when the droplet diameters are larger than the pore diameters. In this respect we note that the precursor droplets are similar to those in the original Millikan oil drop experiment *(19)*, with the same typical charge of –5e. Their motion through air at near atmospheric pressure may be governed largely by electrostatic forces. Thus rather than forming a miniscus, their surface tension is exceeded by capillary action and/or electrostatic forces and complete wetting is achieved; the negatively charged (-5e) liquid droplets are attracted down the trenches, where they produce remarkably uniform dielectric coatings as thin as 40 nm and as thick as 200 nm, depending upon deposition parameters. We emphasize that this uniformity is not obtainable via conventional CVD or sputtering, which produces pooling at the bottom of trenches and inferior step-coverage. The gas phase molecules in CVD deposition are uncharged, so that the same coating mechanism is completely inoperative for structures with such high aspect rations, and incomplete penetration of deposited material results in non-uniform coating.

After deposition, pyrolysis and crystallisation of the precursor results in a coating of the polycrystalline, ferroelectric phase $Sr_{0.8}Bi_{2.2}Ta_2O_9$ on the pore interiors [20]. The surrounding Si substrate can then be removed by wet chemical etching, exposing crystalline SBT tubes. A partially filled Si substrate with pore diameter of *ca.* 2 μm and depth of 80 μm which has been partially filled with SBT is shown in figure 1a. The Si substrate has been selectively etched for a short time to reveal the SBT tubes, which have a uniform wall thickness of *ca.* 200 nm.

A second, more porous substrate with larger pore diameters of *ca.* 3 μm is shown in cross sectional view after complete removal of the host Si walls between pores, figure 1b. This results in a regular array of tubes attached to the host Si matrix only at the tube base. Although these tubes have suffered damage during handling, it is clear that the pores have been filled uniformly to the bottom, a depth of *ca* 100μm. An undamaged array of tubes with diameter *ca.* 800 nm and wall thickness < 100 nm is shown in figure 1c and d. This latter structure can be used as a free-standing photonic array.

**Polymeric Wetting**

The approach consists of wetting of the pore wall of porous templates, either porous alumina or macroporous silicon, by a polymeric precursor containing the metals in the stoichiometric quantities. Lead zirconate titanate ($PbZr_{0.52}Ti_{0.48}O_3$) and barium titanate ($BaTiO_3$) nanotubes were fabricated by infiltrating at room temperature a previously fabricated template [21] by simply bringing in contact the metalorganic precursors and the template. Within a few minutes a thin precursor uniform layer forms on the pore walls from the top down to the bottom of the pores by capillary action. Further investigations showed that the wetting process is so uniform that a complete covering of the whole surface of the pore walls always occurs. As was shown for several polymers, the driving force of the wetting process is the reduction of the surface energy *(22,23)*. Oxide tubes of the appropriate ferroelectric perovskite phase (single-phase) are subsequently obtained by thermal annealing and crystallisation.

Free ferroelectric tubes can be easily obtained by selective etching of the silicon template. Fig. 2a shows straight and smooth $BaTiO_3$ hollow nanotubes with a diameter of about one micron, length of about 50 μm and a wall thickness of 70 nm. The outer diameter can be easily tuned by using different templates. Using nano-porous alumina templates the diameter can range from 50 nm to about 400 nm, while using meso-porous silicon it can vary from 400 nm up to several micrometers *(24)*. The tube length depends on the pore depths and it can be from few micrometers up to more than 100 μm. As shown in Fig. 2b, a single coating produces a tube wall as thin as 50 nm. Increase of the wall thickness is easily achieved by repeated infiltration followed by a low temperature annealing; after the final infiltration the tubes are crystallized by the high temperature annealing. TEM analysis (Fig. 2d) shows that the as-deposited and crystallised tubes of both $BaTiO_3$ and PZT show a tube wall consisting of a crystalline layer sandwiched between two amorphous layers at the silicon- ferroelectric interface and at the internal surface.

**Sol-Gel Dipping**

For ultra-small (20-30 nm) nano-rods and nano-tubes PZT was embedded into porous silicon and alumina matrices by dipping into a precursor (for 3 hours) prepared by a sol-gel method followed by crystallization during thermal annealing. Deposition of sol on the pore walls proceeds as a result of adsorbtion of the sol particles (micelles are positively charged) to membranes pore walls *(24)*. Short immersion times should provide tubules, while longer times yield fibres *(25)* (depending on pore size, concentration, surface tension, shrinkage, etc.). The structure of porous substrates as well as PZT nano-structure was studied by AFM and TEM. It was found that on top of the porous Si nano-structure a PZT layer is formed after dipping. In order to remove the PZT film the surface was either etched by argon beam (for silicon substrate) or cleaned by the solvent rinsing prior to annealing (alumina substrate).

**Hysteresis**

In order to measure the piezoelectric and ferroelectric properties, free standing PZT and BaTiO$_3$ nanotubes were prepared, deposited onto platinum-coated silicon wafer, and subsequently annealed one hour in oxygen at 700°C *(26)*. The ferroelectric properties were measured using scanning force microscopy in the so-called piezoresponse mode *(27,28)*. Individual tubes were probed by a conductive tip and characterized by measuring the local piezoelectric hysteresis. Fig. 3 shows the piezoelectric hysteresis loop obtained on a PZT tube with an outer diameter of 700 nm and wall thickness of 90 nm. The piezoelectric signal is an unambiguous proof the piezoelectricity of the tubes and the hysteresis in the piezoresponse signal is directly associated with the polarization switching and ferroelectric properties of the sample. Moreover, the rectangular shape of the hysteresis loop showing a sharp ferroelectric switching at a coercive voltage of about 2 V is connected with a high quality of ferroelectric material. The effective remanent piezoelectric coefficient is of about 90 pm/V and is comparable with usual values obtained on PZT thin films.

Here we point out that it is difficult to compare the above values to the piezoelectric coefficients of bulk material since the measurement was performed on a tube geometry which has a relatively complicated field distribution and vibrational modes; however, the suggested inference is that the polar axis lies primarily along the nanotube and not through its walls, since the effective d-value is > 10% of the expected value for an oriented film. The single-phase perovskite structure of the tubes was confirmed via X-ray techniques.

**Ultra-Small (20 nm) Devices**

The shape of the nano-tubes depends on the quality of a matrix. In porous silicon the decrease of the pore size down to 10 nm leads to a deterioration of the pore shape and its deviation from cylindrical. Simultaneously, it becomes much more difficult for the precursor to penetrate the pore. Additionally, a thick (>100 nm) PZT film is formed on top of the structure during a dipping, which should be removed by argon beam etching. The thickness of a residual on-top PZT film does not exceed 5 nm and the maximum PZT penetration depth is about 30 nm. Therefore, for a very small-pored silicon substrate a nano-particle rather than nano-tube structure is formed.

For alumina membranes with 20-30 nm pores with high aspect ratio and perfect cylindrical shape can be obtained by an electrochemical etching *(30)*. Since the surface of the pore is smooth, and no constrictions along the pore appear, the penetration depth of a precursor is higher and may reach 100 nm. For PZT/alumina structures the top film is also formed but can be removed by dipping in a solvent. Fig. 5 shows the top AFM image of the PZT/alumina nano-structure (a) and the cross-sectional view (b). After rinsing in a solvent the very top PZT layer is removed not only from the surface, but also form the top of the pores. As a result, a shallow hollow (<10 nm) at the top of each pore is observed. Deep hollows (100 nm) exist for pores in which PZT precursor does not penetrate at all. Therefore, for an alumina substrate, a nano-tube ferroelectric structure is formed.

**Second Harmonic Generation (SHG)**

Optical second harmonic generation (SHG) was used to distinguish between paraelectric and ferroelectric phases of PZT nanolayers embedded in silicon and alumina membranes. Application of this method is based on the difference in the symmetry of the phases: paraelectric phases are centrosymmetric and do not produce SHG in the bulk of the material, while the ferroelectric phase is non-centrosymmetric and produces a very strong SHG signal. For both PZT/Si and PZT/alumina the SHG intensity was increased by more than order of magnitude after annealing, confirming in this way formation of the ferroelectric state in PZT nano-clusters after annealing.

For PZT/Si nano-structures SHG was measured in the reflection geometry at 45 degrees angle of incidence. The laser beam was scanned along the area etched by the argon beam. Figure 5 shows the model of the etched area and the SHG intensity during scanning. The silicon membrane produces small SHG signal whereas the SHG intensity from the film is about the order of magnitude higher. Decreasing the film thickness leads to the decrease of the SHG intensity following by a sharp peak corresponding to the film/nano-structure interface region. Further etching of a nano-structure reduces the signal gradually. A sharp increase of the SHG intensity on top of the PZT/Si nano-structure may arise due to change of the boundary conditions for the ferroelectric particles and due to interference effects as well.

**Summary And Applications**

Given the ferroelectric nature of the deposited phase, the nano-tubes are also piezoelectric, and may have use in a numbers of MEMs applications including ink-jet printing devices, with sub-picoliter droplet size. With further etching it is possible to break the mechanical bond between the tubes and substrate producing free-standing tubes. These tubes may also have interesting potential applications including piezoelectric syringes for drug delivery implants (triggered externally by rf or other stimuli). In addition to the application in Si DRAM trenching, these processes may allow fabrication of devices for mass storage analogous to the IBM millipede *(31)*, and tunable [2D] photonics.

**Acknowledgements:**
Work supported by Cambridge-MIT grant CMI-001, the DFG under WE496/19 and WE 2637/1, INTAS 75-2002, CRDF and RF Ministry of Education VZ-010-0, the

Volkswagen Stiftung, and by Samco Corp. We thank T. Leedham, U. Gösele and J. Schilling, O. Tsuji and T. Tatsuta for helpful discussions. We thank S. Schweizer, S. Matthias, J. Choi and H. Masuda for the preparation of templates.

**Materials and Methods**

Liquid precursor (from Inorgtech) for chemical solution deposition consisted of a 0.1 M solution of Sr-, Bi-, Ta-ethylhexanoates in toluene, with appropriate ratios of Sr, Bi and Ta to give the resulting stoichiometry $Sr_{0.8}Bi_{2.2}Ta_2O_9$. PZT 9906 Polymer and BATIO 9101 Polymer (*Chemat Technology*) were used for pore wetting. PZT sol-gel precursor was prepared via anodic dissolution of Ti-Zr in $CH_3OC_2H_4OH$ and dehydration of $Pb(CH_3COO)_2 \cdot 3H_2O$ with $(CH_3CO)_2O$. Crystallisation by thermal annealing was carried out in air at 650-700°C (30 mins, PZT), 800 °C (1 hr, SBT) and 850 °C (1 hr, $BaTiO_3$). Wet etching was carried out in an aqueous solutions of 15 vol% HF / 50 vol% $HNO_3$ or 20 wt % KOH solution at 90°C.

Optical (SHG) measurements were performed with the use of a 100-fs Ti-sapphire amplified laser at the fundamental wavelength of 800 nm and with a femtosecond optical parametric amplifier at the fundamental wavelength in the range of 670 - 520 nm. The repetition rate was 1 kHz; pulse energy, 0.05 mJ; spot diameter, 100 μm. SHG radiation after filtering by monochromator and colour filters was detected by a PMT and a lock-in amplifier.

**References:**

1. D. M. Antonelli, Y. J. Ying, *Angew. Chem. (Int. Ed. Engl.)* **34**, 2014 (1995).

2. D. Shi, W. J. v. Ooij, *Appl. Phys. Lett.*, **78**, 1234 (2001).

3. V. V. Pokropivny, 201$^{st}$ Electrochemical Soc. Meeting abstract and *Physica* **C351**, 71 (2001) claims piezoelectric nanotubes of boron nitride deposited on top of SiC cores.

4. Piezoelectric ink-jet printer sleeves: K. Herzog, E. Kattner (Siemens AG), US Patent Number 4504845 (12 March 1985).

5. Ferroelectric ink-jet printers: S. Sakamaki, M. Aizawa, M. Toki, Y. Yamada, Y. Takahashi, US Patent Number 20010412 (15 Nov 2001).

6. Tunable photonics with ferroelectrics: J. Sajeev, K. Busch, US Patent Number US2002074537 (20 June 2002).

7. Deep trenching of DRAMs with ferroelectric capacitors: B. Gnade, P. Kirlin, S. Summerfelt (Texas Inst.), US Patent Number US6033919 (7 March 2000).

8. J. Averdung, M. Droescher, A. Greiner, J. H. Wendorff, German Patent Number DE10023456 (2 January 2001).

9. International Technology Roadmap for Semiconductors (ITRS) 2002 (available at http://public.itrs.net/Files/2002Update/Home.pdf).

10. B.A. Hernandez, K.S. Chang, E.R. Fisher, *Chem. Mater.*, **14**, 481 (2002).

11. G. R. Fox, *J. Mater. Sci. Lett.*, **14** 1496 (1995).

12. Infineon Corp. (private communication).

13. Tatsuya Shimoda (private communication).

14. L. D. MacMillan *et al.*, *Integr. Ferroelectrics* **2**, 351 (1992).

15. M. Huffmann, *Integr. Ferroelectrics* **10**, 39 (1995).

16. N. Solayappan, L.D. McMillan, C.A. Araujo, R. Grant, *Integr. Ferroelectrics* **18**, 127 (1997).

17. J. Schilling *et el.*, *Appl. Phys. Lett.*, **78**, 1180 (2001).

18. S. Ottow, V. Lehmann, H. Foell, *Appl. Phys.* **A63**, 153 (1996).

19. R. A. Millikan, *Science* **32**, 436 (1910).

20. F. D. Morrison *et al.*, *Microelectron. Eng.*, in press (Proc. Int. Conf. Mater. Res. Soc., Xi'an, China, June 2002).

21. R. B. Wehrspohn, J. Schilling, *MRS Bull.* **26**, 623 (2001).

22. V. Lehmann, *J. Electrochem. Soc.* **140**, 2836 (1993).

23. M. Steinhart *et al.*, *Science* **296**, 1997 (2002).

24. E. D. Mishina *et al.*, *J. Exp. Theor. Phys.* **95**, 502 (2002).

25. J.C. Hulteen, C.R. Martin in *Nanoparticles and Nanostructured Films*, J.H. Fendler Ed. (Wiley-VCH, Weinhem, Germany, 1998).

26. M. Tomkiewiez, S. Kelly, in *Nanoparticles and Nanostructured Films*, J.H. Fendler Ed. (Wiley-VCH, Weinhem, Germany, 1998).

27. This high temperature treatment allows converting of the amorphous layer into the perovskite ferroelectric phase and releasing defects and residual stress after etching process.

28. O. Auciello *et al.*, *MRS Bull.* **23,** 33 (1998).

29. An atomic force microscope provided with a conductive tip and a lock-in detection system is used to measure the piezoelectric vibrations generated by the sample itself via converse piezoelectric effect by applying an ac voltage across the sample.

30. H. Masuda, T. Mizuno, N. Baba, T. Ohmori , *J. Electroanal. Chem.* 368, 333 (1994).

31. P. Vettiger *et al.*, *IBM J Res. Dev.* **44,** 323 (2000).

**Figure captions**

**Figure 1.** SEM micrograph indicating a plan view of a regular array of SBT tubes in host silicon substrate with diameter *ca.* 2 µm and wall thickness *ca.* 200 nm (a). Larger diameter SBT tubes in cross sectional view indicating coating to bottom of pore (b). Micrograph of free standing array of tubes with diameter *ca.* 800 nm (c) and wall thickness < 100 nm (d).

**Figure 2.** Scanning electron microscopy image of (a) a bundle of free $BaTiO_3$ tubes, (b) of an opening end, and (c) of capped tip of $BaTiO_3$ nanotubes. (d) Cross-section transmission electron microscopy images of PZT nanotubes in silicon template.

**Figure 3.** Piezoelectric hysteresis loop of an individual PZT tube measured by piezoresponse AFM.

**Figure 4.** AFM image of PZT nanostructure embedded into porous alumina (a) and probe signal in cross section (b) indicating filled (•) and empty (o) pores.

**Figure 5.** Model of the PZT/porous Si structure after etching (a) and SHG intensity along the etched area and porous membrane (b).

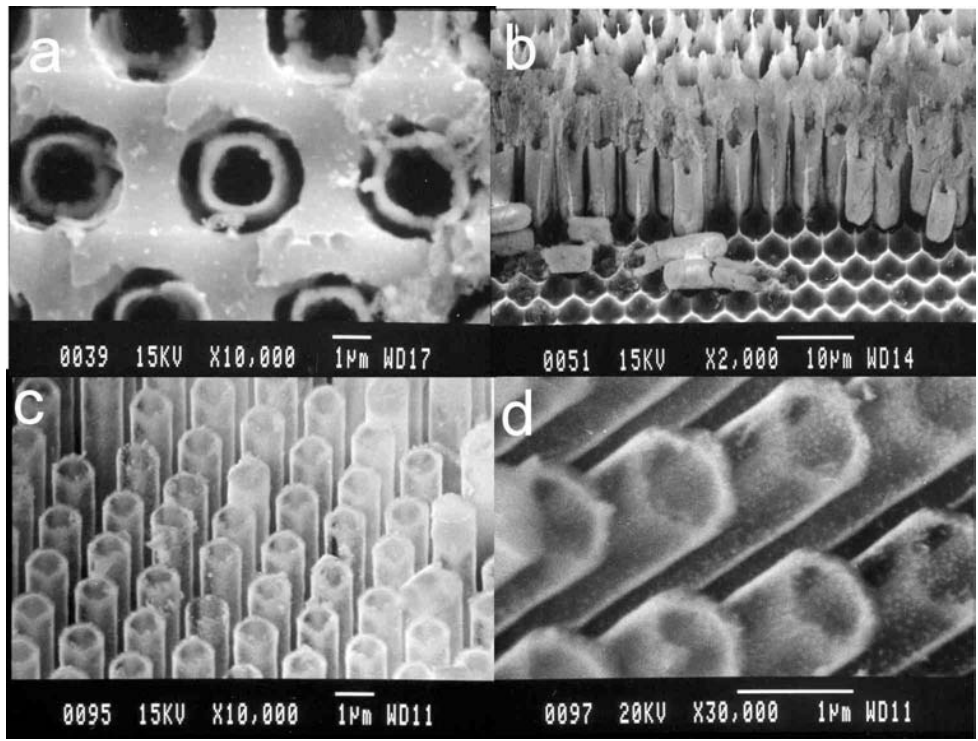

Fig 1

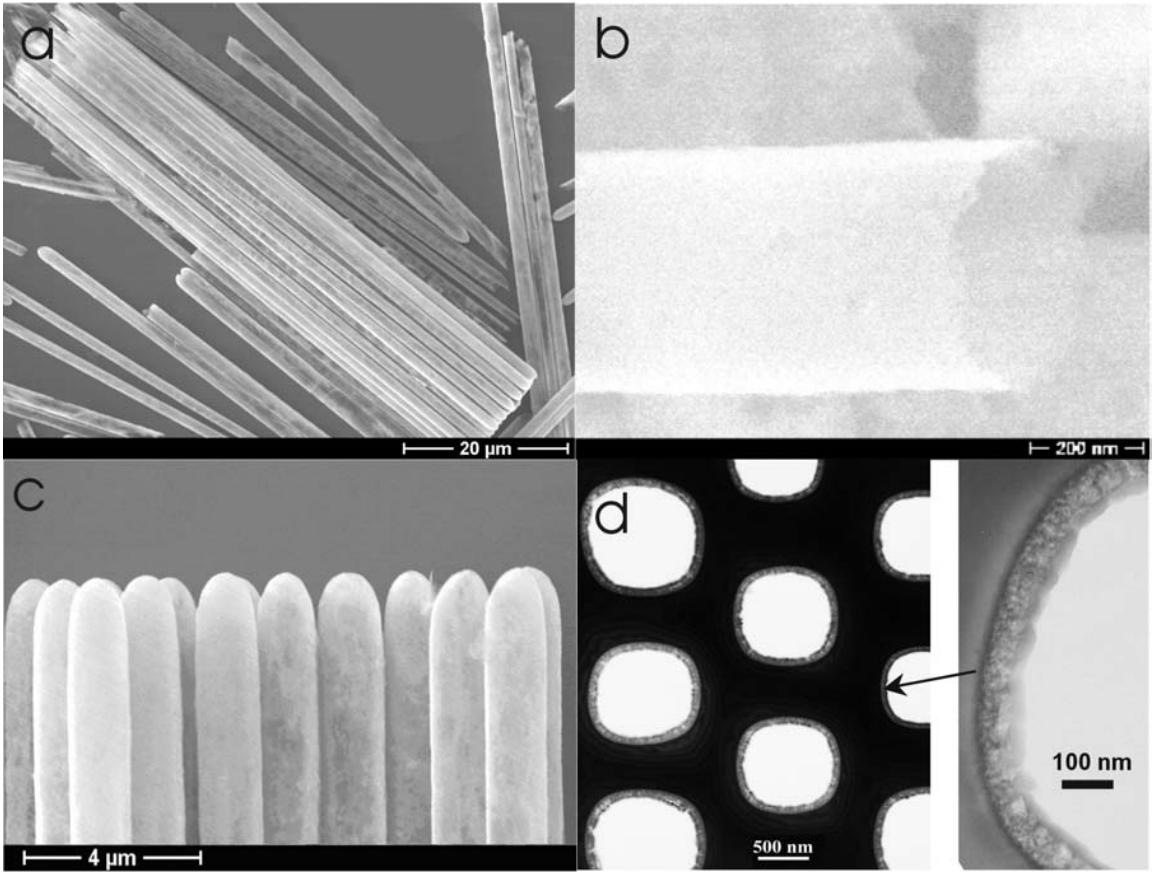

Fig 2

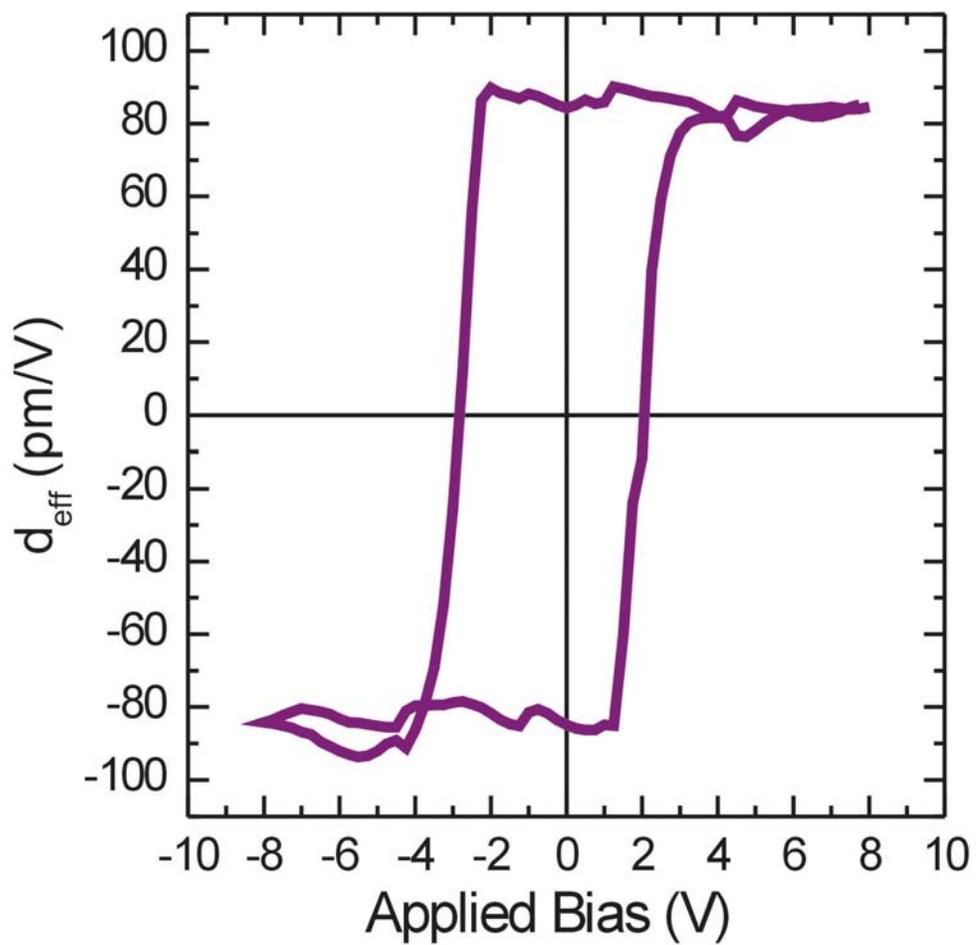

Fig 3

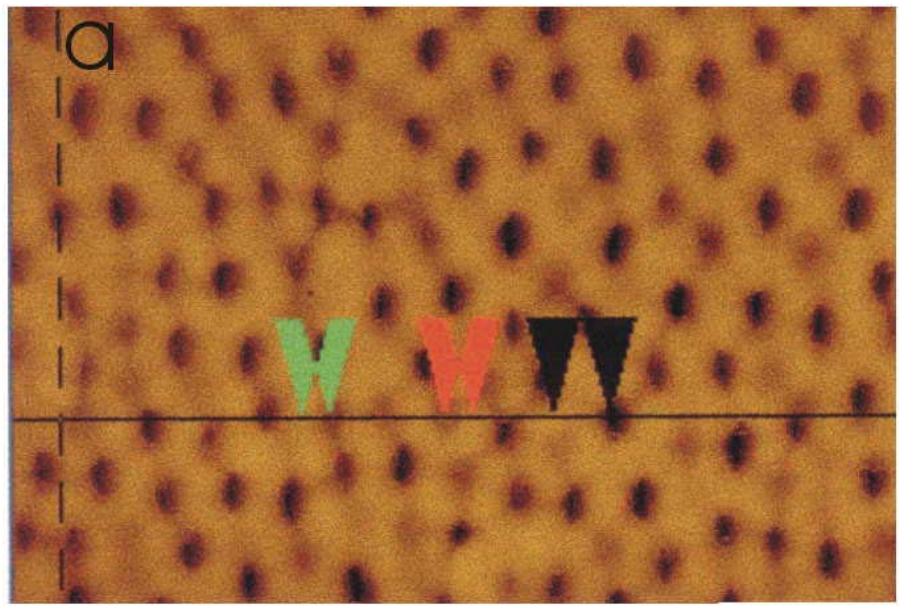
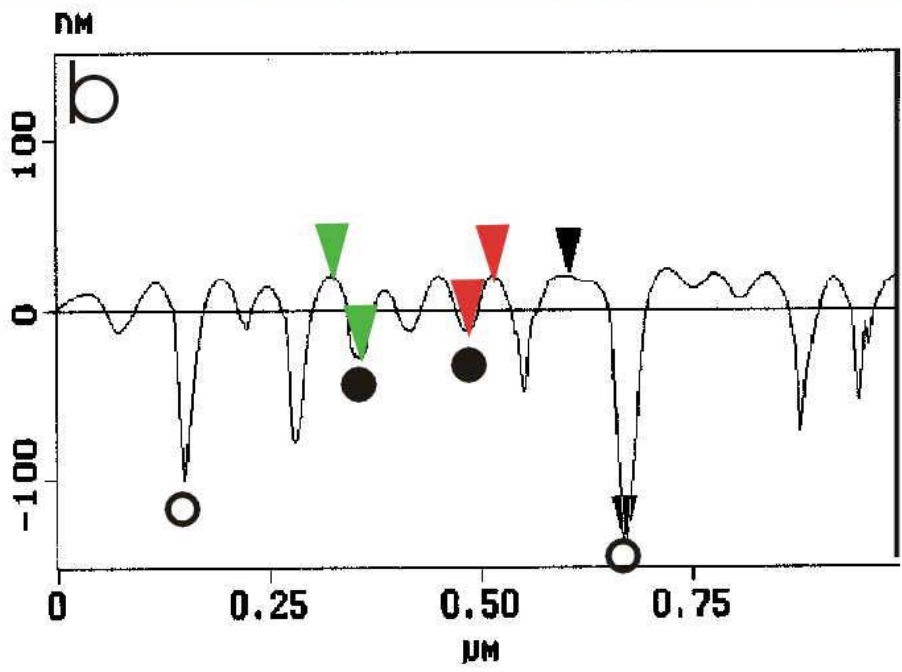

Fig 4

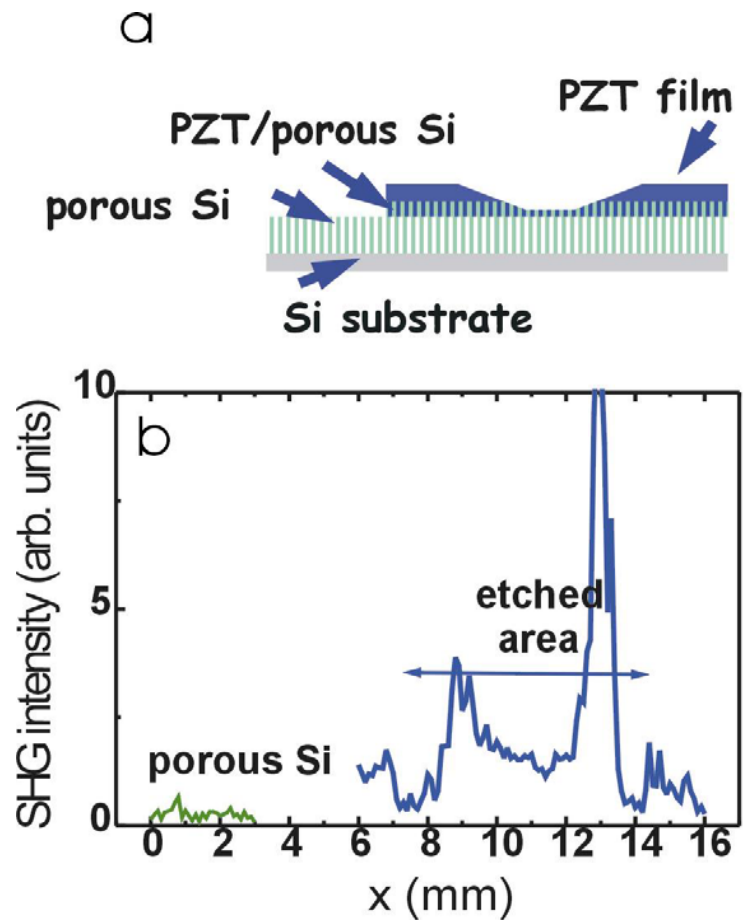

Fig 5